\documentclass[oneside,a4paper,12pt,english]{article}
\usepackage[italian]{babel}
\usepackage[utf8]{inputenc}
\usepackage[T1]{fontenc}
\usepackage{amssymb}
\usepackage{amsbsy}
\usepackage{amsmath}
\usepackage[pdftex]{graphicx}
\usepackage{fancyhdr}
\usepackage{hyperref}
\usepackage{enumitem}
\usepackage{chemfig}
\usepackage{wrapfig}
\begin{document}


\thispagestyle{plain}
\begin{center}
    \Large
    \textbf{Dal tramonto all'alba}
        
    \vspace{0.4cm}
    \large
Breve ricognizione sulla fisica italiana nel primo trentennio del XIX secolo
        
    \vspace{0.4cm}
    \textbf{Benedetto P. Casu}
       
    \vspace{0.9cm}
    \textbf{Abstract}
\end{center}
Questo lavoro presenta una ricognizione dello stato della ricerca fisica nella penisola italiana nel trentennio 1800 - 1830. Vengono seguite tre direttive metodologiche: la scelta di concentrarsi sul Regno di Sardegna e sui territori del Lombardo - Veneto, per il ruolo giocato nel presente (e nel futuro) politico e scientifico della penisola; un approccio quantitativo per l'analisi dell'articolazione della comunità scientifica basato sulla teoria delle reti; infine, un'attenzione dedicata alle "personalità minori", minimizzando il ruolo dei "giganti" del periodo (Volta, Avogadro, Lagrange, per esempio) per concentrarsi sullo sviluppo della comunità scientifica autoctona.

\newpage

\tableofcontents


\listoffigures

\newpage

\section{Introduzione. Scelta metodologica}

\begin{scriptsize}
\begin{flushright}
« L'aver avuto un passato è sempre stato, nella storia delle istituzioni italiane, un motivo sufficiente per domandare un pezzetto d'avvenire che poi si sarebbe sommato, accrescendolo, al passato, e avrebbe fornito i titoli per richiedere una maggiore quota di futuro »\\
Luigi Besana \emph{L'Accademia scientifica nel periodo della sua formazione e costituzione}, Torino, 1978
\end{flushright}
\end{scriptsize}

Lo scopo di questo articolo è di offrire una breve ricognizione dello stato della fisica italiana nel primo trentennio del XIX secolo. Cogliendo lo spunto metodologico di Pietro Redondi \cite{redondiCulturaScienzaDall1978}, seguiremo un'analisi basata su tre ambiti: un'area geografica, che include il Regno di Sardegna e il Lombardo-Veneto, ossia due entità che giocarono un ruolo importante negli sviluppi della comunità scientifica nella penisola italiana tra la fine del XVIII secolo e il 1861; un ambito biografico, con un tentativo di analisi quantitativa dell'articolazione interna alla comunità scientifica dell'epoca basato sulla teoria delle reti e infine un taglio generale che escluda le personalità più in vista (Spallanzani, Volta, Avogadro, Lagrange, per illustrare il calibro). L'ultima assunzione è forse la più discutibile e l'analizzerò in dettaglio successivamente. Le prime due necessitano comunque di una precisazione. Ho "tagliato via", con l'accetta, importanti aree della penisola: la Repubblica di San Marco, lo Stato pontificio, il Granducato di Toscana, il Regno delle due Sicilie, i ducati di Parma e Piacenza, il Ducato di Modena, la Repubblica di Lucca. Queste sono aree che, sottolineiamolo, hanno apportato un contributo alla storia della comunità scientifica, un contributo che è rimasto nel tessuto sociale e tecnico dell'Italia unita, in termini di personale impegnato nella ricerca, accademie, associazioni, istituti di ricerca (molta parte del granulare sistema universitario italiano ha sede in queste zone, per esempio). Ma la natura di questo lavoro impone una scelta chiara e definita ad un ambito di ricerca che sarebbe altrimenti vasto quanto una Nazione intera ed attraverserebbe decine e centinaia di contributi. Per quanto riguarda invece l'ambito "biografico", è mia intenzione evitare di cadere in una discussione di storia della scienza dove manca la "storia" ed invece è preponderante la "scienza": dunque una storia di idee. Intendo invece seguire la da un lato l'impostazione epistemologica tratteggiata da Boris Hessen nel Congresso di Londra del 1930\footnote{B. Hessen, \emph{Science at the cross Road}, Frank Coss \& Co. 1971, pg, 166-167} in cui venne messa in luce la stretta interdipendenza delle condizioni ambientali - economiche, politiche, sociali - con l'elaborazione scientifica; dall'altro, una trattazione che sappia tessere una cornice narrativa in grado di presentare il periodo storico come autonomo - nei limiti della concezione di "autonomia" per una periodizzazione che è comunque arbitraria ultima analisi e tratta di fenomeni che si estendono, per cause ed effetti, ben al di là della finestra temporale individuata - per favorirne la ricezione e rendere più facile coglierne l'importanza.
\\
L'idea di effettuare un'operazione di sottrazione, spostando sul margine le figure che invece i contemporanei e una buona parte della storiografia successiva hanno invece messo in primo piano, coglie le mosse da un'intuizione dello stesso Redondi che cita il Gramsci dei "corpi catalitici":
\begin{quote}
[\ldots] Si tratterebbe insomma di uno studio dei «corpi catalitici» nel campo storico-politico italiano, elementi catalitici, che non lasciano traccia di sé ma hanno avuto una insostituibile e necessaria funzione strumentale nella creazione del nuovo organismo storico\cite{gramsciQuaderniDalCarcere2014}.
\end{quote}
Questo permette di mettere in luce il contributo di uno strato molto ampio che la storiografia posteriore ha esaltato, spesso in funzione teleologica, avente come obiettivo l'unificazione nazionale e la conseguente costruzione di una Scienza nazionale. Quella dell''unità è un'aspirazione certamente presente nel periodo pre-Risorgimentale, che nell'intellettualità diffusa della penisola italica data indietro fino a Dante, Petrarca e Boccaccio; non può però diventare l'unica chiave di lettura acriticamente acquisita. Per usare le parole dello stesso Redondi:
\begin{quote}
[\ldots] non c'è alcun dubbio che gli scienziati italiani del secolo XIX guardavano la scienza con una radicata «coscienza del sapere nazionale», come disse Cesare Correnti nel 1872. Non neghiamo infatti l'esistenza di una ricerca di identità nazionale che si fece avvertire nei diversi scienziati sparsi negli Stati italiani e in esilio. Ciò che neghiamo è che quella consapevolezza implichi per lo storico della scienza in Italia una nozione non problematica, un dato acquisito. In altre parole, neghiamo una prospettiva risorgimentista che ricavi dal motto «scienza e nazionale» degli scienziati ottocenteschi una categoria storiografica di italianità capace di riunire sotto di sè le idee scientifiche presenti in Italia.
\end{quote}
Aggiungiamo che questo è tanto più problematico per un periodo come quello in esame, che rischia di essere schiacciato dalla presenza di alcuni ingombranti \emph{vicini}: alle sue spalle, la Rivoluzione francese e l'ascesa di Napoleone, sul piano storico, e la potente opera di sistematizzazione della meccanica razionale sul piano scientifico, esemplificata da formulazioni come il cosiddetto \emph{demone di Laplace}; dall'altro, in avanti, il Risorgimento, che per l'Italia ha rappresentato una stagione fecondissima di impegno civile e crescita della coscienza nazionale, ma che il fascismo ha, in seguito, in una prospettiva nazionalista, eretto a canone per l'intero XIX secolo, rileggendo tutti gli eventi da quell'osservatorio privilegiato. Il primo trentennio del XIX secolo, invece, necessita di un approccio "leggero", dove si possa portare in luce anche le sfumature più tenui. Il periodo in esame si presta infatti ad una molteplicità di analisi: prendiamo per esempio il primo intervento diretto francese nella penisola nel 1796. Dovremmo considerarlo una invasione o piuttosto la risposta ad una guerra "esterna" portata avanti contro un nemico anch'esso esterno e spesso in rotta con le "aspirazioni nazionali" dei patrioti italiani, per definirlo un "intervento amico"? Le Repubbliche sorte per volontà dei Francesi sono un esempio di "colonialismo" e di oppressione, oppure un occasione di autogoverno? Lo sviluppo scientifico degli Stati italiani è stato "autonomo" oppure si è trattato di uno sviluppo di temi e situazioni importate dall'esterno, dovuta all'arretratezza congenita delle comunità scientifiche locali? Risulta evidente che non esista una risposta unica ma che ogni questione è attraversata da contraddizioni, spesso laceranti, che hanno influenzato con forza un'epoca che si pone a cavallo di importanti snodi della storia europea e mondiale. Una continua tensione fra diversi centri di forza che rappresentano un punto di interesse da preservare nell'analisi del periodo.

\newpage

\section{La situazione italiana alla fine dell'ultima decade del XVIII secolo}

\begin{scriptsize}
\begin{flushright}
« All who served the revolution have plowed the sea »\\
Simon Bolivar, towards the end of his life, about 1830.
\end{flushright}
\end{scriptsize}

\begin{wrapfigure}{r}{0.5 \textwidth}
\centering
\includegraphics[width=0.5 \textwidth]{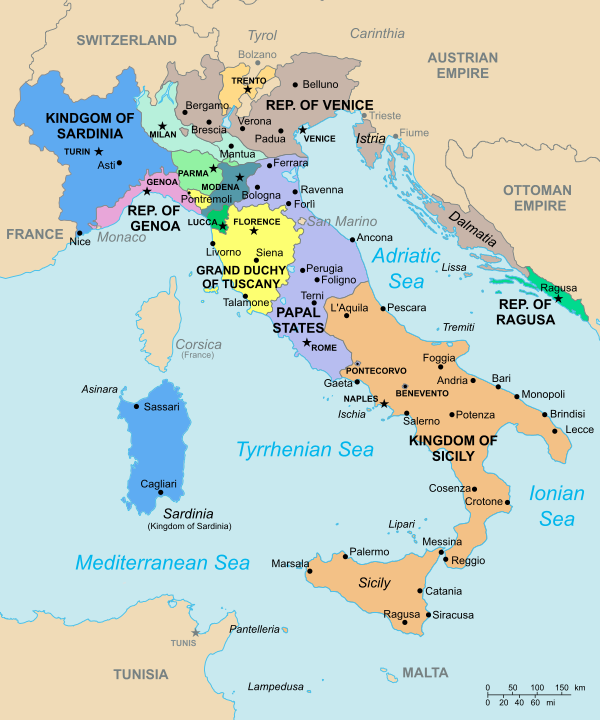}
\caption{Situazione politica della penisola italiana alla vigilia del primo intervento francese dell'aprile 1796}
\end{wrapfigure}

La mappa mostra la frammentazione politica della penisola italiana nel 1796, prima della spedizione francese comandata dal generale Bonaparte. Oltre alla differenza di governi ed istituzioni, dobbiamo pensare alle differenze in termini di linguaggio, leggi, valute e sistemi di pesi e misure, che complicavano enormemente gli scambi sia fisici che di conoscenza.
\\
Nel 1796 il Direttorio francese, nella sua lotta contro gli Asburgo, decise di approntare una spedizione in Italia con lo scopo di assicurarsi la tranquillità al confine sud-est. Tra la primavera del 1796 e l'aprile del 1797 Napoleone, al comando dell'Armata d'Italia, sconfisse la coalizione tra il Regno di Sardegna e l'Impero asburgico, arrivando pericolosamente vicino a Vienna (a soli 250 km, per la precisione\footnote{\cite{bantiEtaContemporaneaDalle2009}, pg. 95}). Con il trattato di Campoformio (o Campoformido nel dialetto veneto dell'epoca) venne sancita la devoluzione della maggior parte dei territori della ex-Repubblica di San Marco all'Austria in cambio della cessione del Lombardo-Veneto e della fine dell'ingerenza di Casa Asburgo nello scacchiere italiano (ricordiamo che gli Asburgo erano direttamente coinvolti nelle vicende toscane, in cui la casa regnante era un ramo cadetto degli Asburgo stessi, gli Asburgo - Lorena), direttamente e indirettamente. I territori così "liberati" vennero riorganizzati nelle cosiddette "Repubbliche sorelle", entità politico-amministrative autonome in cui era comunque evidente la funzione di tutela della Francia (in quello che potremmo modernamente considerare un "protettorato" francese).
\\
Le cosiddette "repubbliche sorelle" sono un gruppo di repubbliche instaurate con il supporto dell'esercito francese di occupazione nel corso del triennio 1796 - 1799. Il Direttorio francese, e Napoleone in qualità di comandante dell'Armata francese in Italia, cercò di instaurare governi amici che sollevassero i francesi dall'amministrazione diretta delle terre occupate, rendendo più semplice e agevole la permanenza su suolo straniero. Le repubbliche non furono mai realmente indipendenti e si appoggiarono al corpo di occupazione francese per i compiti di polizia, esercito e spesso anche per compiti amministrativi come l'esazione delle tasse. Nonostante ciò, le repubbliche del 1796/1799 rappresentano il primo tentativo di unitarietà nel governo della penisola italica.

\begin{wrapfigure}{r}{0.5 \textwidth}
\centering
\includegraphics[width=0.5 \textwidth]{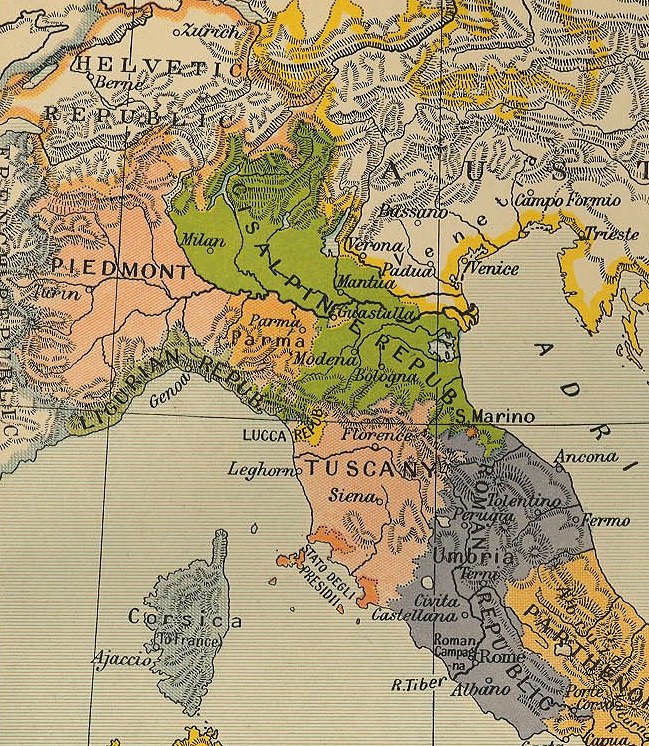}
\caption{Repubbliche dell'Italia settentrionale, 1799 \cite{CisalpineRepublic}}
\end{wrapfigure}

\section{Il panorama geografico e politico}

\subsection{Il Regno di Sardegna}
Qui è doverosa una ulteriore precisazione di carattere geografico ma che avrà importanti ripercussioni anche sul futuro di una parte importante dell'Italia. Se infatti ho utilizzato qui la dizione di "Regno di Sardegna", è stato per intendere il Piemonte, che del Regno di Sardegna rappresentava il cuore dell'economia e delle istituzioni politiche e culturali, laddove la Sardegna stava in un rapporto di subordinazione che si esprimeva anche in campo scientifico, mancando nell'isola istituzioni e personale impegnato in ricerche autonome. Tale situazione rende difficoltoso, in questa sede, dare menzione di contributi che non appaiono in evidenza nella storiografia consultata, ma che pur dobbiamo supporre siano esistiti, in qualche forma. Nel proseguo mi limiterò quindi a delineare i caratteri economici e scientifici del Piemonte. 
\\
Nel trentennio preso in esame, la situazione economica del Piemonte è caratterizzata da una molteplicità di fattori frenanti il suo sviluppo che possiamo raggruppare in tre classi:
\begin{itemize}
\item la \emph{dominazione francese}, che pur dotando la regione di istituzioni e politiche ricalcate su quelle francesi, spesso all'avanguardia confrontate con quelle sabaude, ne asserviva le risorse ad obiettivi stranieri, non favorendo una ricaduta positiva sul territorio delle azioni intraprese;
\item la \emph{concentrazione di capitali} nelle mani di un ceto nobile interessato alla rendita ed alla speculazione fondiaria (per l'agricoltura) e in commercianti interessati alla vendita di materie prime e semilavorati nei mercati europei avanzati; 
\item la \emph{politica protezionista} dello Stato centrale, che proteggeva soprattutto la produzione agricola da merci a basso costo provenienti dalla Russia, ma che non incentivava uno sfruttamento razionale delle terre, che si trovavano quindi ad essere coltivate con strumenti e tecniche non efficienti.
\end{itemize}
Lo spartiacque della Restaurazione, con il suo colpire ogni movimento riformista, non incentivò un cambiamento in queste dinamiche, sebbene lo stesso Prospero Balbo cercasse di trovare un punto di equilibrio tra le istanze (che potremmo definire) \emph{di mercato} e un controllo statale che assicurasse la stabilità sociale (anche se soprattutto da una prospettiva aristocratica e non a favore delle classi subalterne)\footnote{cfr. M. Ciardi, \emph{La fine dei privilegi}, L. Olschki editore, 1995, Firenze, pg. 133}. I moti del 1821 riacutizzarono le paure della Corte, che fece retrocedere i tentativi di Balbo, così che nel 1830 si poteva definire ancora il Piemonte <<uno  Stato ad involuto indirizzo agricolo>>\footnote{\emph{Ibidem}, pg. 135}.

\subsubsection{Il Privilegio. Il ruolo del brevetto agli esordi dello sviluppo industriale piemontese}
Il brevetto è un protagonista importante dello sviluppo industriale, ma la sua forma attuale discende da un'evoluzione storica complessa che ha attraversato differenti fasi in differenti aree geografiche. In questa sede, la sua importanza deriva dal ruolo che giocò nello sviluppo della rivoluzione industriale in Piemonte. Esso rappresenta uno snodo importante nel rapporto tra scienza, tecnica ed evoluzione dell'imprenditoria, con ricadute in ogni settore. In Inghilterra esso era regolato fin dal 1623 dallo <<Statuto dei Monopoli>> emanato da Giacomo II nel 1623 e si basava su una semplice richiesta di registrazione\footnote{\emph{Ib.}, pg. 146} da parte del proponente. In Francia, e questa sarà la linea del Piemonte nel periodo in esame, esso era legato a due condizioni: novità e utilità. La prassi era quella di sottomettere la richiesta ad un parere dell'Accademia delle Scienze, che aveva le competenze e il prestigio necessari per l'istruttoria. 
\\
Alla fine del XVIII secolo, però, la pressione dell'innovazione divenne tale che l'Accademia non fu più in grado di reggere il ritmo delle proposte e nel 1791 questo esame venne rimosso\footnote{\emph{Ib.}, pg. 147}. A seguito dell'occupazione francese, questa modifica si propagò anche nel Piemonte annesso, fino a quando Vittorio Emanuele I, ritornato in possesso del Regno, il 21 Maggio 1814 restaurò la legislazione precedente. Nonostante le difficoltà dello sviluppo piemontese, nel corso degli anni successivi il numero di richieste aumentò e si dovette mettere mano ad una riforma, che portò all'approvazione di una nuova legge il 28 Febbraio 1826, \emph{Regie patenti, colle quali sono stabilite nuove regole per la concessione delle Privative industriali}. Tale legge riformava la concessione del Privilegio in senso "moderno", legandolo ad un esame delle caratteristiche tecniche, affidato all'Accademia delle Scienze di Torino. La caratteristica importante ai fini di questa analisi lega la nuova legge alla consuetudine del Regno di Sardegna di porre sotto Privilegio, e spesso finanziare, quelle invenzioni che avessero una ricaduta positiva nel ramo d'industria indicato. Tale principio era inteso così importante che si legava la concessione alla dimostrazione che l'attività del proponente fosse in buona salute e già avviata da tempo, per <<andare di riparo di alcuni abusi, come per es. quando il ramo d'industria privilegiato fosse dal concessionario tenuto in attività  per uno o per pochi anni, e non più; nel qual caso il privilegio resterebbe senza effetto con danno dell'industria comune\footnote{Regie Patenti, 28 Febbario 1826, art. 10, in M. Ciardi, \emph{La fine dei privilegi}, L. Olschki editore, 1995, Firenze, pg. 151}>>. Con questa legge l'Accademia delle Scienze di Torino si dotava inoltre di un regolamento per disciplinare l'analisi tecnica delle richieste; Amedeo Avogadro, in qualità di membro dell'Accademia, si interessò personalmente di molti dei procedimenti, la cui analisi delle carte ci permette di trarre molte utili informazioni sul rapporto tra scienza, tecnica e sviluppo industriale come veniva inteso all'epoca. Nel corso della prima relazione a cui prese parte, riguardante delle macchine per la filatura della seta e di altri tessuti, la commissione riporta:
\begin{quote}
<<I deputati non dubitano che l'uso di queste macchine abbrevierà e diminuirà di molto la man d'opera per le operazioni che ne formano l'oggetto, le quali sin qui generalmente, e in Piemonte, e per quanto essi credono anche nell'estero, si fanno almeno in gran parte a mano \ldots [Questa riduzione permette di] diminuire con gran vantaggio del pubblico, il prezzo delle materie preparate, come già si è sperimentato dall'uso delle macchine da pochi anni introdotte per la filatura del cotone>>\footnote{\emph{Ib.}, pg. 147}.
\end{quote}
La connessione tra introduzione della macchina e riduzione del prezzo è già evidente e definita.
\\
L'impianto della Regie Patenti trae origine da una più antica attidudine dello Stato sabaudo nei confronti di quella che potremmo definire con un anacronismo "politica industriale", legata al ruolo dello Stato nella gestione dell'economia. Nel corso del XVIII secolo si venne delineando un organismo, il Consiglio di Commercio, un organismo consultivo in materia economica, con poteri di vigilanza sull'intero Regno\footnote{Luisa Dolza, \emph{<<A vantaggio di pochi e danno evidente di molti>>: Amedeo Avogadro, la tecnica e i privilegi in piemonte nella prima metà dell'Ottocento}, pg. 2, in \cite{ciardiFisicoSublimeAmedeo2007a}}. Le fonti mostrano come il ruolo di questo Consiglio fosse di analisi ed indirizzo delle innovazioni proposte a tutela della redditività dell'industria ma anche della stabilità sociale ad essa collegata. Il privilegio era quindi utilizzato come leva per favorire lo sviluppo di un settore, ma veniva accordato in un quadro di garanzie per le realtà locali coinvolte (soprattutto nel caso di un proponente estero, in linea con la politica protezionista allora egemone nel pensiero economico europeo). In questo era molto importante, per esempio, la formazione delle maestranze, che veniva esplicitamente richiesta come condizione per la concessione\footnote{Luisa Dolza, \emph{<<A vantaggio di pochi e danno evidente di molti>>: Amedeo Avogadro, la tecnica e i privilegi in piemonte nella prima metà dell'Ottocento}, pg. 4, in \cite{ciardiFisicoSublimeAmedeo2007a}}. Spesso inoltre non era tanto il privilegio ad essere accordato, ma finanziamenti per ricompensare l'inventore e permettere la commercializzazione dell'invenzione. Vi era anche un'attenta analisi del ruolo distorsivo del monopolio nell'influenzare lo sviluppo di un ramo d'industria\footnote{\emph{Id.}, pg. 3}.
\\
Con la Restaurazione e "l'effervescenza dovuta alla rivoluzione industriale", che aumenta notevolmente il carico di lavoro del Consiglio e la complessità dell'analisi dei manufatti, si delinea una situazione di tensione tra Consiglio di Commercio ed Accademia delle Scienze, situazione che verrà risolta con la succitata legge del 1826, che trasferirà la competenza interamente nelle mani dell'Accademia\footnote{Luisa Dolza, \emph{<<A vantaggio di pochi e danno evidente di molti>>: Amedeo Avogadro, la tecnica e i privilegi in piemonte nella prima metà dell'Ottocento}, pg. 6, in \cite{ciardiFisicoSublimeAmedeo2007a}}.

\subsection{Il Lombardo-Veneto}

\subsubsection{Le riviste scientifiche}
Una componente importante dello sviluppo scientifico del Lombardo-Veneto durante questo trentennio passa attraverso le riviste scientifiche pubblicate in loco. Fra questi, all'indomani della Restaurazione (1815/1816) ebbero una certa importanza il "Giornale di fisica, chimica, storia naturale, medicina ed arti dè professori Brugnatelli, Brunacci e Configliachi, compilato dal Dottore Gaspare Brugnatelli", su cui Avogadro pubblicherà una serie di lavori, e la "Biblioteca Italiana". Quest'ultima era stata creata su impulso del governo austriaco per mostrare pubblicamente la sua volontà di riconciliazione, ma fin da subito mostrò invece la reale intenzione di mantenere la discussione intellettuale nell'alveo di una politica filo-austriaca che risultò indigeribile per una certa parte dell'intellettualità lombarda\footnote{cfr. M. Ciardi, \emph{La fine dei privilegi}, L. Olschki editore, 1995, Firenze, pg. 62}. Dall'opposizione a questa esperienza nacque "Il Conciliatore", pubblicato a cavallo tra il 1818 e il 1819, diretto da Silvio Pellico, tra i quali collaborò anche il Romagnosi. Al Conciliatore parteciperà attivamente l'astronomo reale sabaudo Giovanni Plana, professore di matematica presso l'Università di Torino e importante punto di riferimento matematico per Avogadro.
La "Biblioteca" nasceva con un forte intento divulgativo, proponendosi di recensire le nuove pubblicazioni di carattere medico, fisico e letterario ma anche contributi originali di memorie tecniche e scientifiche\footnote{\emph{Ib.}, pg. 64}. Si noti come la fraseologia impiegata riporti spesso l'utilizzo dei termini "Italia", "Italiani" e si riportino contributi di giornali e riviste pubblicate in altri luoghi della penisola, non direttamente sotto controllo austriaco\footnote{\emph{Ib.}, pg. 64}. 
%
 
\section{Il panorama epistemologico}

\subsection{Il Piemonte}

\subsubsection{Empirismo e ricerca teoretica nello stato Sabaudo}
In questo clima la ricezione del programma di ricerca di Laplace all'interno dei confini italiani e segnatamente nel Regno di Sardegna, che per vicende storiche e posizione geografiche meglio degli altri si prestava ad una stretta connessione con lo sviluppo della scienza dell'ingombrante Vicino, assume la forma dell'evoluzione della cattedra di Fisica Sublime all'Università di Torino. "Sublime" è un aggettivo che indicava, in campo matematico, la branca del calcolo integrale e differenziale\footnote{M. Ciardi, \emph{La fine dei privilegi}, L. Olschki editore, 1995, Firenze, pg. 71}. In Fisica, esso denotava quella che chiamiamo "meccanica razionale", cioè la matematizzazione della fisica che assumeva in quel momento storico la funzione di universalizzazione dell'approccio newtoniano e rappresentava un "ponte" tra scienze teoriche e sperimentali. In quest'ottica la disciplina si caratterizzava come centrale per lo sviluppo del Piemonte e così era considerata dai protagonisti dell'epoca, Prospero Balbo e Amedeo Avogadro. Nei progetti di riforma che seguirono la Restaurazione, a cui un personaggio come Balbo non aderì mai fino in fondo, pur provenendo da posizioni monarchiche e non rivoluzionarie, alle scienze toccava un posto centrale per mantenere lo Stato (piemontese) all'avanguardia nella competizione economica e militare internazionale\footnote{\emph{Ib.}, pg. 72}.

\subsection{Il Lombardo-Veneto}

\subsubsection{La ricezione di Newton e Leibniz nella cultura scientifica milanese}
Il clima culturale del Lombardo-Veneto nei primi decenni del XIX secolo affonda le sue radici nell'inedito sviluppo che la regione seppe avere sotto il dominio austriaco, dove spinte politiche ed economiche seppero saldarsi per creare le condizioni di uno sviluppo massiccio\footnote{P. Redondi, \emph{Cultura e scienza dall'illuminismo al positivismo}, in AA.VV., \emph{Storia d'Italia. Annali}, vol III, pg. 686}. L'elaborazione epistemologica risente in grande misura del tentativo di accordare le conseguenze più durature degli approcci filosofici al problema della scienza moderna di Newton e Leibnitz. Sul primo infatti si basa la decisa opera di \emph{meccanizzazione} della fisica, dove cioè ogni problema veniva inserito in un quadro di interazioni reciproche la cui intensità si basava su leggi di potenza (l'inverso del quadrato della distanza in analogia con l'attrazione gravitazionale), con la conseguenza che il movimento della materia (delle sue componenti) usurpava il centro dell'analisi rispetto alle sue qualità, come era stato invece nella fisica aristotelica\footnote{\emph{Ib.}, pg. 689}. La cultura illuministica aveva poi interpretato questa conseguenza in chiave decisamente antimetafisica, limitando l'analisi delle cause del movimento per concentrarsi invece soltanto su di esso, operazione esplicitamente annunciata da D'Alembert:
\begin{quote}
<<Conseguentemente a questa riflessione ho per così dire distolto la vista dalle cause motrici per non esaminare altro che il solo movimento che esse producono\footnote{J. B. Le Rond D'Alembert, \emph{Traitè de dynamique}, 2nd ed. Paris, 1758, pp. XVI sg., in P. Redondi, \emph{Cultura e scienza dall'illuminismo al positivismo}, in AA.VV., \emph{Storia d'Italia. Annali}, vol III, pg. 690}.>>
\end{quote}
Questo non poteva non porre problemi di ordine religioso in una società in cui gli organismi ecclesiastici avevano un peso notevole, e questo accadeva anche nella scienza\footnote{A titolo d'esempio, tra i maggiori protagonisti del dibattito epistemologico della seconda metà del XVIII secolo a Milano troviamo Giuseppe Boscovich e Paolo Frisi, rispettivamente un gesuita e un barnabita}. Con l'intento di recuperare una cornice trascendente veniva utilizzato Leibnitz, con il suo rifiuto della teoria del discreto della materia e l'accento sulla sua infinita divisibilità\footnote{P. Redondi, \emph{Cultura e scienza dall'illuminismo al positivismo}, in AA.VV., \emph{Storia d'Italia. Annali}, vol III, pg. 689}. Di conseguenza, non era \emph{l'impulso} ad essere la quantità principale in gioco ma la \emph{forza}, cosa che lascia spazio ad un'ipotesi di unificazione in senso anche metafisico\footnote{\emph{Ib.}, pg. 691}. Si ricreava così la possibilità per la metafisica di giocare un ruolo positivo interno all'epistemologia, in una maniera che permetteva di recuperare anche lo stesso Galileo, nonostante i famosi suoi problemi. Dulcis in fundo, una teoria del continuo poteva essere geometrizzata e sottoposta ad un'operazione di ordinamento e gerarchizzazione analoga all'assiomizzazione della geometria, che permetteva quindi un approccio deduttivo basato su una base empirica\footnote{\emph{Ib.}, pg. 689}.
\\
Sul versante più decisamente antimetafisico, sempre prendendo le mosse da Newton, si muovevano coloro i quali che aderivano più decisamente allo sviluppo in senso fisco-matematico della meccanica, in accordo con la tendenza maggioritaria nella comunità scientifica francese\footnote{\emph{Ib.}, pg. 693}. Questo si esplicava nell'utilizzo dell'analisi matematica e dei suoi procedimenti all'interno di un'elaborazione che includesse il dato sperimentale senza però assegnarli una posizione decisamente superiore. In questo clima vennero quindi accolte le sollecitazioni all'utilizzo della matematica in maniera rigorosa e produttiva, estendendo gli ambiti in cui l'analisi poteva essere utilizzata per risolvere problemi. Qui si trova però anche il limite di un approccio che, pur non negando l'importanza della teoria, aveva comunque, per ragioni anche extrascientifiche - il riformismo teresiano e l'impulso allo sviluppo produttivo - la necessità di trovare una giustificazione nella sua capacità di \emph{risolvere problemi}. Per questo troviamo un esito essenzialmente "operativista" alla scienza nel Lombardo-Veneto, che nella dialettica con il dato "concreto" non seppe uscire da una riflessione binaria empirismo/idealismo\footnote{\emph{Ib.}, pg. 697}. Il pragmatismo con cui la cultura scientifica lombardo-veneta affrontò il secolo permise comunque al pensiero scientifico di circolare e farsi portatore di un impegno civile che avrebbe poi dato i suoi frutti nella stagione del Risorgimento\footnote{\emph{Ib.}, pg. 698}.

\subsubsection{Lo sviluppo del dibattito epistemologico nel primo trentennio del XIX secolo}
Il pensiero di Ambrogio Fusinieri si inserisce nella polemica ottocentesca tra Boscovich e Frisi esprimendo una certa sintesi delle posizioni antimeccanicistiche e vitaliste che attraversavano la critica alla fisica-matematica ed al programma di ricerca di Laplace. Il nocciolo fisico delle sue teorie si basa su un dinamismo interno alla materia, generato da una forza repulsiva che specchia l'esistenza di una forza attrattiva di carattere gravitazionale fra le molecole\footnote{P. Redondi, \emph{Cultura e scienza dall'illuminismo al positivismo}, in AA.VV., \emph{Storia d'Italia. Annali}, vol III, pg. 721}. Tale forza repulsiva, che si osservava quando la materia si trovata in "forma attenuta", cioè riscaldata e/o sotto forma di lamine sottili e pellicole (come le bolle di sapone, fenomeno già studiato da Newton a suo tempo), portava alla conclusione di una forza interna responsabile di un processo di allontanamento infinito delle componenti della materia, un processo di suddivisione infinitesimale che trovava in Leibniz una giustificazione matematica\footnote{\emph{Ib.}, pg. 722}. Questa ipotesi/osservazione era comune ad una serie di fenomeni che riguardavano calore, elettricità e trasformazioni fisiche e giustificava l'idea di una fondamentale \emph{unitarietà} dei fenomeni. Qui sta l'importanza di Fusinieri nel panorama scientifico dell'epoca: la sua ricerca di un principio unitario, che esprimeva una tensione epistemologica e teologica a una spiegazione generale dei fenomeni fisici, e la sua concezione empirista della ricerca scientifica. Entrambi i tratti - unitarietà e sperimentalismo - lo pongono come interprete della sua epoca, segnata da fermenti romantici e positivisti in critica con la concezione illuministica e razionalista ben esemplificata da Laplace e dai "francesi" (con cui Fusinieri battaglierà a lungo, nella persona di Poisson, in merito alla fisica dei fluidi\footnote{\emph{Ib.}, pg. 723}). Tale apparente modernità lo poneva all'avanguardia rispetto ad una fisica-matematica che, con il suo approccio matematizzante ed il suo distacco dall'esperienza, non riusciva a dare conto delle relazioni tra fenomeni diversi e si trovava a postulare apparenti contraddizioni (l'esistenza degli atomi come centri del moto posti a distanze molto grandi rispetto all'entità delle forze gravitazionali supposte esistenti fra loro, il vuoto interno alla materia su cui si propagavano forse senza mediatori), ma la sua epistemologia tutta fondata su un empirismo "semplice" non andava oltre, come criterio metodologico, al richiamo all'esperienza come giudice della validità "a priori" di una ipotesi scientifica. Mancava un'analisi del ruolo della matematica come strumento predittivo delle relazioni tra entità, cosa che il positivismo in elaborazione non mancava di dare\footnote{\emph{Ib.}, pg. 726}. Egli scrive:
\begin{quote}
[\ldots] In luogo di abbandonarsi a tali finzioni [l'esistenza di fluidi imponderabili come il calorico, il vuoto fra gli atomi, la loro natura indivisibile], le quali non possono che impedire i veri progressi della scienza, se deve limitarsi a quanto i fenomeni presentano, scegliendo i più semplici e i più generali e classificando i fatti. In ciò solo può consistere il vero fondamento del sapere\footnote{A. Fusinieri, \emph{Riflessioni generali contro la teoria degli atomi e contro quella degli imponderabili}, in \emph{Annali delle scienze del regno lombardo-veneto}, vol. V, 1835, pp.gg. 149-153, in \footnote{\emph{Id.}, pg. 726}}.
\end{quote}
L'esito, per lo più infelice, di queste elaborazioni avrebbe condotto ad un certo ristagno nella comunità scientifica italiana fino alla vigilia del Risorgimento, che avrebbe rivitalizzato il dibattito attraverso una partecipazione in prima persona del personale scientifico nella penisola italica.


\newpage

\section{Il panorama biografico. Un abbozzo di analisi delle relazioni tra i soggetti in esame}

Questa sezione è un tentativo di analisi quantitativa del network fra gli scienziati nel periodo in esame creato dalla comune appartenenza a due istituzioni, la Reale Accademia delle Scienze di Torino e l'Accademia Nazionale delle Scienze detta dei XL. 
L'immagine di apertura rappresenta il totale delle interazioni reciproche fra i membri delle due istituzioni nel periodo in esame. 

\begin{figure}[h]
\centering
\includegraphics[width= \textwidth]{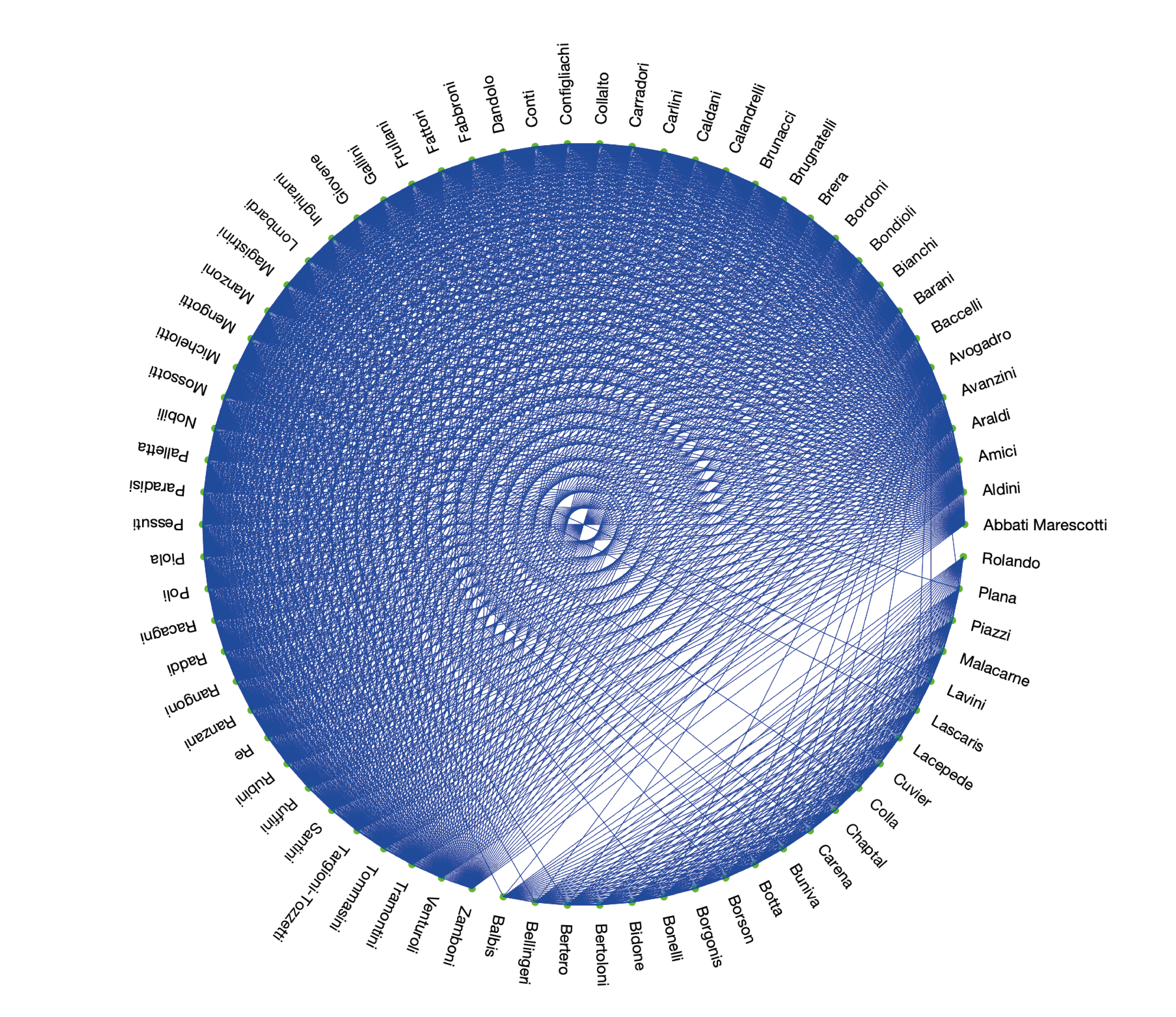}
\caption{La comunità scientifica ristretta agli/alle appartenenti alle due istituzioni nel primo trentennio del XIX secolo}
\end{figure}

Nell'immagine qui sotto ho invece ridotto il numero delle connessioni fra i membri interni alla stessa Accademia all'arco di cerchio che connette i punti, in modo da concentrarsi sulle interconnessioni fra le due comunità. 

\begin{figure}[h]
\centering
\includegraphics[width= \textwidth]{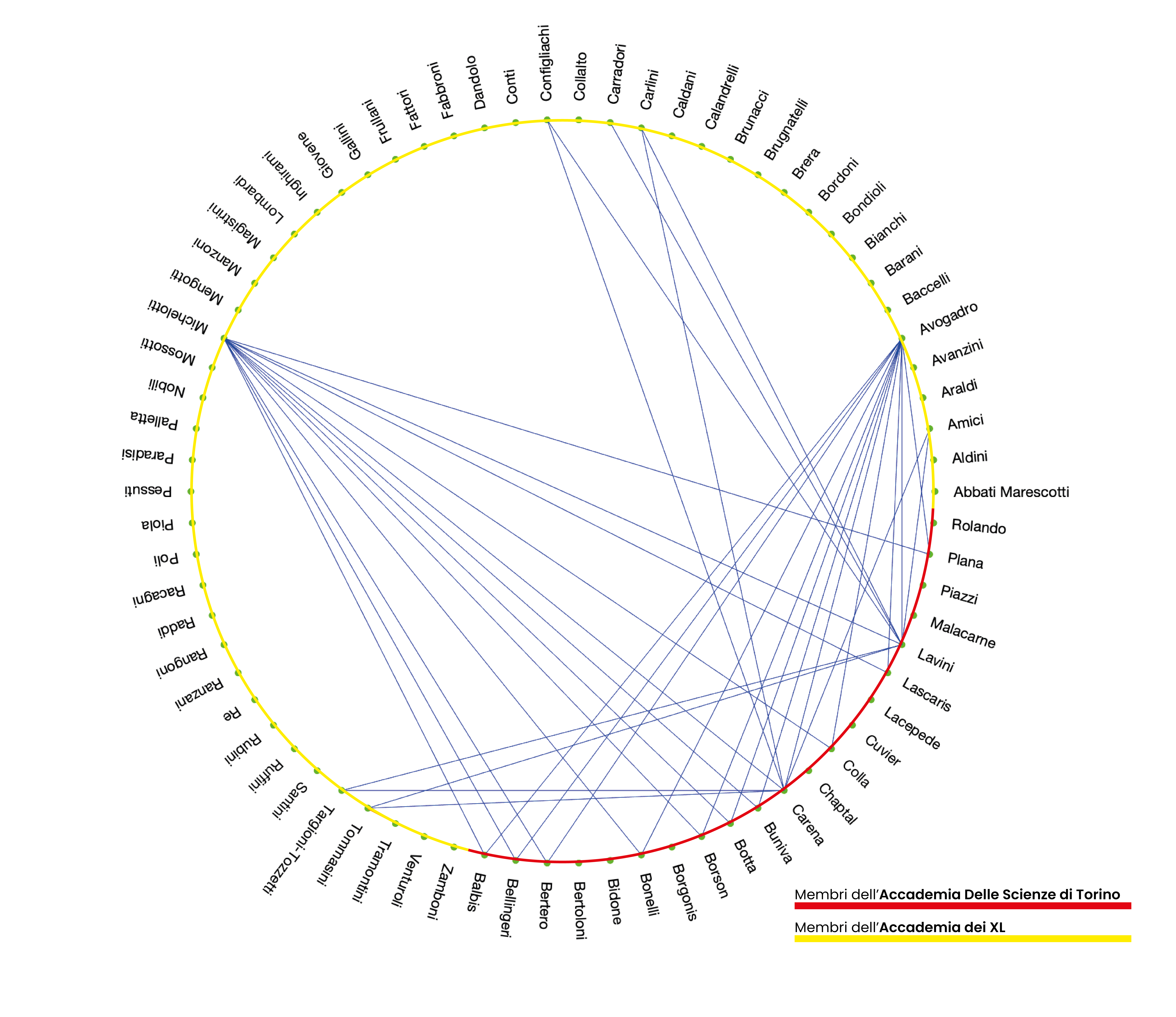}
\caption{Connessioni fra scienziati mediate dall'appartenenza alla medesima istituzione}
\end{figure}

I dati sono stati recuperati dall'elenco dei membri dell'Accademia dei XL su Wikipedia\footnote{\url{https://it.wikipedia.org/wiki/Soci_dell\%27Accademia_nazionale_delle_scienze}, consultato per l'ultima volta in data 14 giugno 2021}, estrapolandone i membri iscritti tra il 1800 e il 1830, mentre per la Reale Accademia delle Scienze di Torino sono stati ripresi dal volume \emph{Il primo secolo della R. Accademia delle scienze di Torino. Notizie storiche e bibliografiche. (1783-1883)}, edito nel 1883 tra gli Atti dell'Accademia e digitalizzato nel 2011 dalla University of Toronto\footnote{Accessibile all'indirizzo: \url{https://teca.accademiadellescienze.it/book/ilprimosecolodel00acca} grazie all'Internet Archive}.
\\
Si tratta di una rete composta da 74 membri, la gran parte (53) iscritta all'Accademia dei XL, che aveva nello Statuto l'indicazione a non tenere in considerazione la provenienza geografica dei suoi iscritti, laddove l'Accademia di Torino discriminava fra "soci nazionali" (cioè del Regno di Sardegna) e "soci stranieri" (come sono, nel nostro caso,  Pietro Paoli e Franz Xaver von Zach, rispettivamente di Pisa e Bratislava).
\\
Nel proseguo, ho assunto che la partecipazione alla stessa istituzione presupponga la presenza di un link fra i soggetti e la presenza contemporanea in due (o più) istituzioni costituisca un link tra le due comunità. La rete è assunta come indiretta, per rappresentare lo scambio fra soggetti ed istituzioni nei due sensi. Un piccolo caveat - fra i molti che si possono immaginare nell'affrontare una trattazione di questo tipo: ho cercato di restringere i soggetti ai soli membri delle classi di Scienze fisiche (fisica e chimica); non sempre questo restituisce risultati in linea con le aspettative contemporanee, dove gli ambiti di ricerca sono chiaramente delineati. Non era raro trovare, durante l'epoca studiata, molti medici e quelli che definiremmo "scienziati naturali" interessarsi attivamente a ricerche più propriamente di fisica e chimica. 
\\
Come si può osservare dalla figura, sono pochi i membri che condividono la doppia iscrizione, segno di una regionalizzazione della comunità che restituisce un network dove pochi nodi hanno la responsabilità di funzionare come \emph{bridge} fra i sottoinsiemi. Il grado medio <\emph{k}> è pari a 3,2, mentre si osserva che gli \emph{hub} Avogadro, Michelotti, Carena e Lavini hanno un grado molto maggiore (rispettivamente 12, 12, 4, 4). 

\newpage

La figura successiva mostra una struttura ad albero della rete. Si noti la struttura "a lobi" del network, tenuto insieme da soli quattro hub, di diversa importanza.

\begin{figure}[h]
\centering
\includegraphics[width= \textwidth]{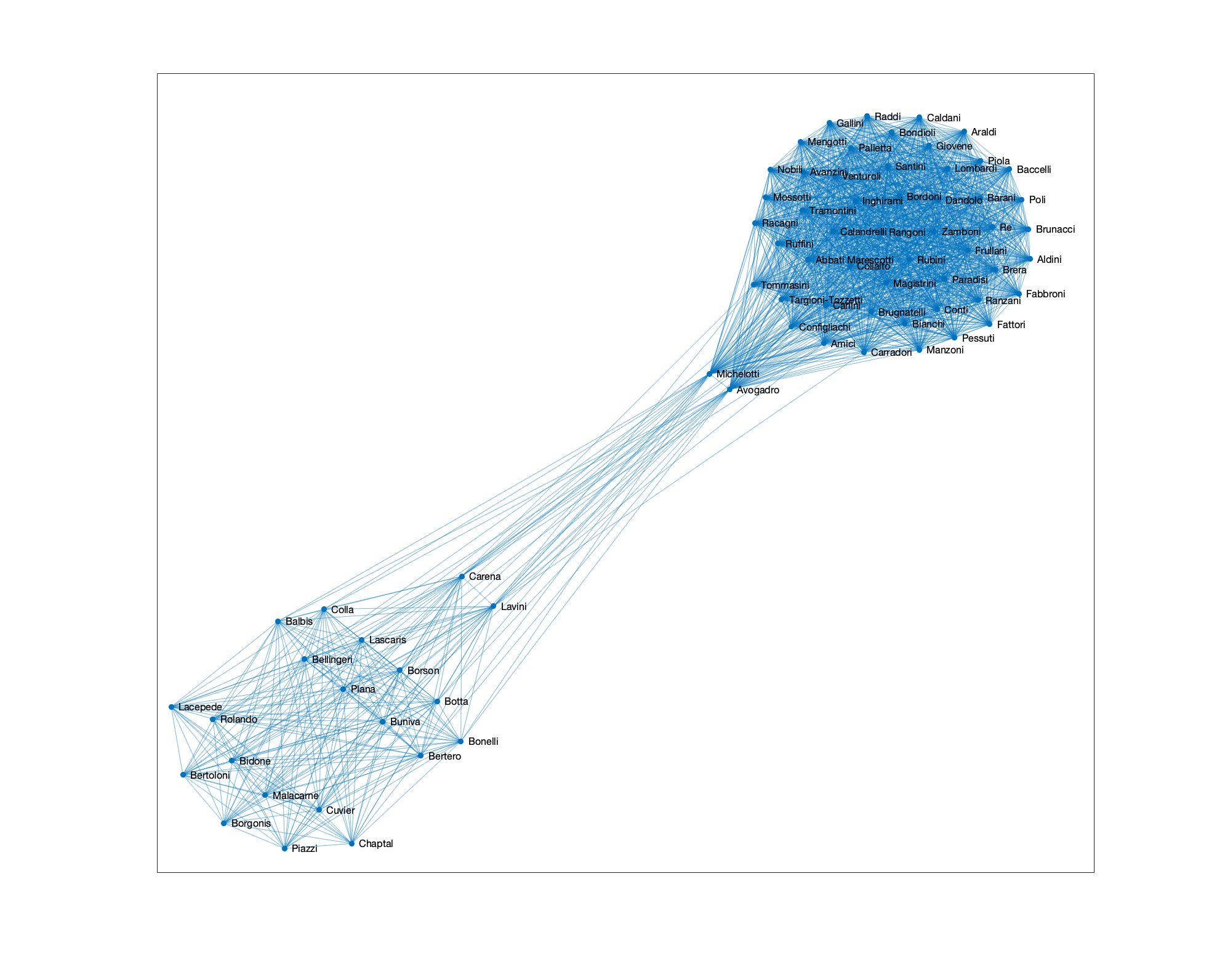}
\caption{Grafo ad albero della rete}
\end{figure}

Nel seguito ho invece posto a zero il collegamento fra i membri della stessa istituzione per concentrarci sul sottoinsieme dei bridge (l'uso dei colori distingue fra le due comunità).

\begin{figure}[h]
\centering
\includegraphics[width= \textwidth]{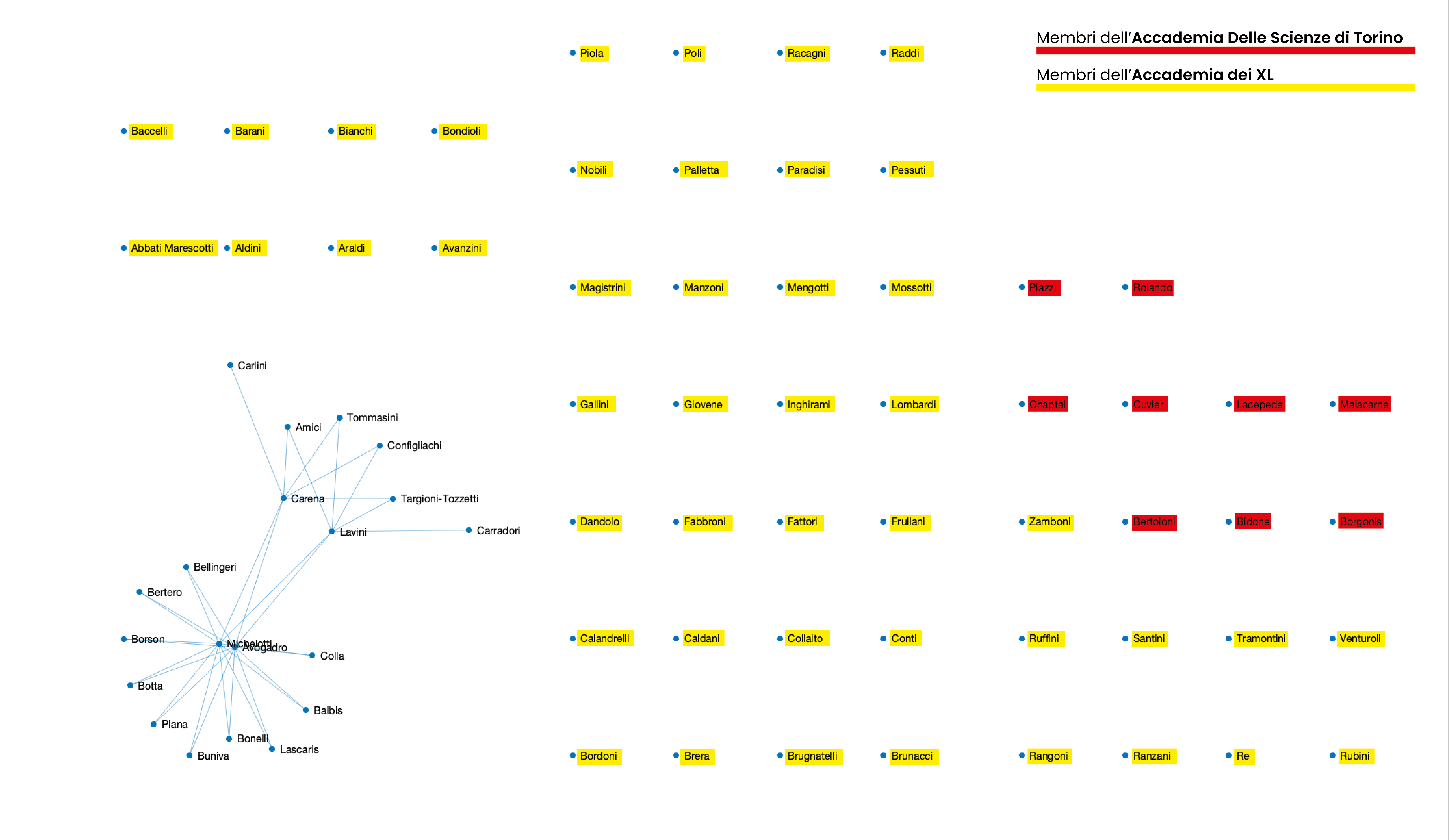}
\caption{Grafo ad albero della rete, con l'isolamento del sottoinsieme più connesso}
\end{figure}

La fragilità della comunità scientifica del periodo rispecchia la profonda regionalizzazione allora imperante e le difficoltà poste dall'ambiente circostante nel portare avanti una ricerca organica. La presenza delle Accademie costituiva al tempo stesso una \textbf{opportunità} di connessione ma un \textbf{limite} alla condivisione, nella misura in cui all'interno delle Accademie si aveva la possibilità di essere aggiornati sui lavori altrui (con i limiti che i rituali e la natura dei membri ponevano, in un'epoca in cui lo "scienziato di professione" stava solo iniziando a prendere forma) ma il numero e la diffusione locale delle istituzioni portava ad una frammentarietà che inibiva la formazione di gruppi di lavoro "internazionali".
\\
\newpage

L'ultima figura mostra come un evento esterno porti al rafforzamento della struttura: la Prima Riunione degli Scienziati Italiani a Pisa nel 1839. Ho ricavato i dati dall'elenco dei partecipanti alla Riunione raccolti nel volume \emph{Prima riunione dè naturalisti, medici ed altri scienziati italiani tenuta in Pisa nell'ottobre 1839}, stampato nell'ottobre 1839 dalla Tipografia Nistri e digitalizzato dal Museo Galileo\footnote{\url{https://bibdig.museogalileo.it/Teca/Viewer?an=301346}, url consultato l'ultima volta il 14 giugno 2021}. 

\begin{figure}[h]
\centering
\includegraphics[width= \textwidth]{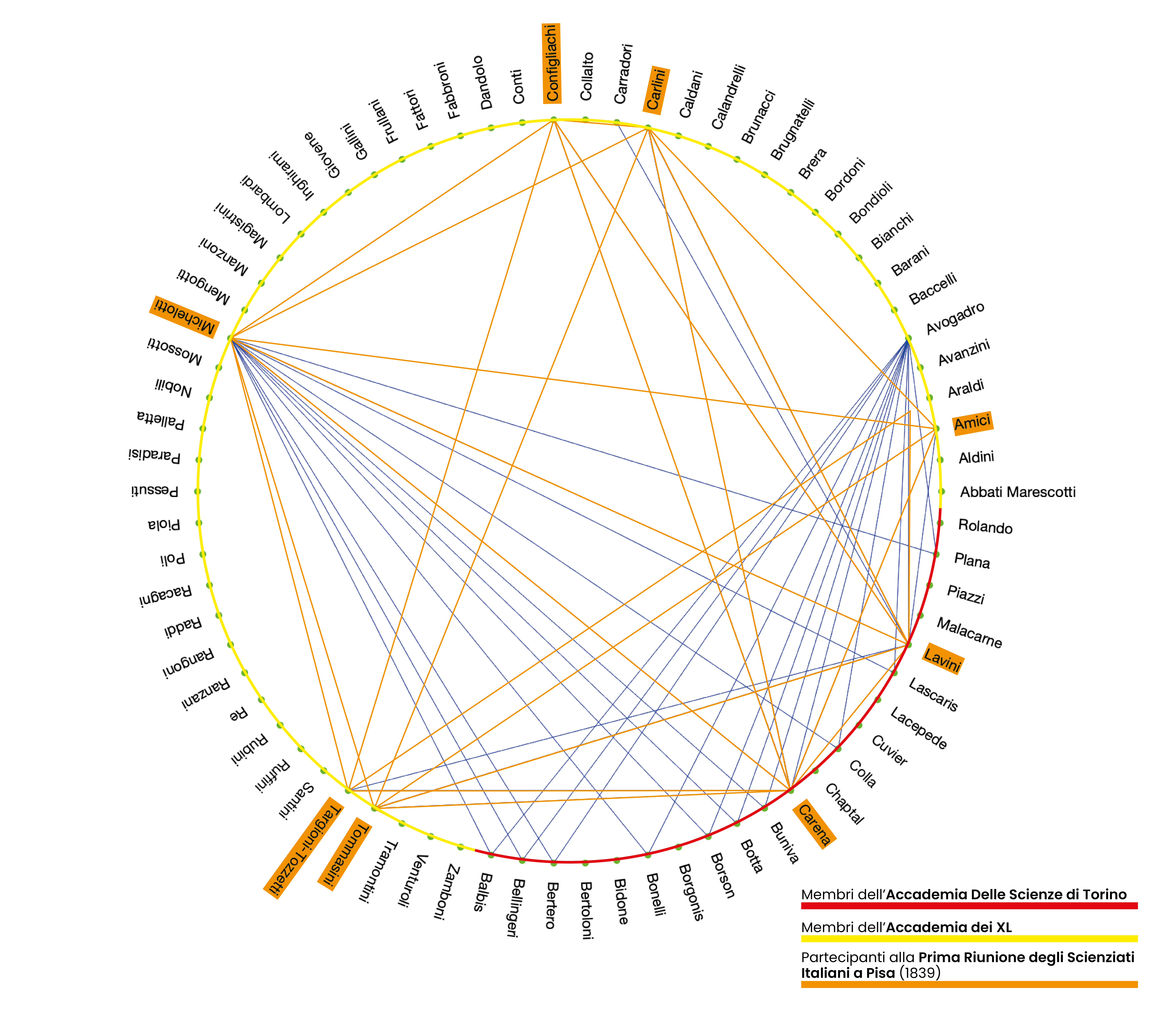}
\caption{I cambiamenti nella comunità con l'introduzione delle Riunioni degli scienziati italiani nel 1839}
\end{figure}

Si può osservare un irrobustimento degli hub già presenti (Avogadro non risulta presente alla Riunione), con Michelotti che passa da un grado di 12 a 19 e Carena da 4 a 10 e la formazione di nuovi, come Configliachi (per altro Presidente della Sezione di Scienze Naturali della Riunione), Carlini, Lavini, Amici, Tommasini e Targioni - Tozzetti (che un certo peso ebbe anche nella definizione della nascente letteratura romantica).


\section{Conclusioni}

Il periodo in esame permette alcune riflessioni sullo sviluppo scientifico della penisola italiana e sulle sue ripercussioni sull'attualità. Le profonde differenze regionali, in termini di sviluppo interno e di connessione con l'esterno (Regno Unito e Francia in primis, ma nel volgere di qualche decennio anche l'area di lingua germanica avrebbe fatto il suo ingresso unificato sul proscenio internazionale) hanno giocato un ruolo preponderante nel determinare gli orientamenti della comunità scientifica "protoitaliana". Questo insieme di circostanze ambientali si riflettono tutt'ora nella varietà e vitalità delle istituzioni scientifiche: basti pensare che entrambe le Accademie citate sono ancora operanti, così come le decine di università che esistevano all'epoca. Il rapporto dialettico fra centri e periferie continua ad essere un fattore determinante anche nella ricaduta della ricerca scientifica sui settori produttivi dell'economia, con una differenza marcata fra Nord e Sud del Paese ed una differenziazione notevole all'interno degli stessi distretti produttivi\footnote{\url{https://www.istat.it/it/files/2020/07/Livelli-di-istruzione-e-ritorni-occupazionali.pdf}, url consultato in data 14 giugno 2021}. La "debolezza" della comunità scientifica, brevemente messa in luce nelle ultime pagine, ispira una serie di questioni e richieste di approfondimenti sul suo stato attuale. 
\\
Seppur centrale nello sviluppo della comunità scientifica, dunque, rimane la netta sensazione che l'epoca in esame sia stata letta con una certa difficoltà dalla storiografia di parte. Ho accennato brevemente alle difficoltà incontrate nel reperire fonti secondarie con un fuoco diretto sull'argomento; tale difficoltà è accentuata dalle letture successive del periodo \emph{in funzione} dei periodi successivi: il Risorgimento, soprattutto, e la lettura che del Risorgimento ne fece il fascismo. Come ricorda il prof. Barbero in una conferenza del 15 maggio 2021 a Lucca\footnote{\url{https://www.youtube.com/watch?v=umUFW0n5LFA}, url consultato il 14 giugno 2021} riferendosi all'esperienza delle Repubbliche sorelle filofrancesi, prevale nella storiografia la necessità di riconnettere lo sviluppo dell'Italia unita con il passato mitico piuttosto che con il vicino prossimo; ecco quindi il recupero del Regno longobardo d'Italia e il vagheggiare del ritorno dell'Impero sui "colli fatali" di Roma, piuttosto che il riferimento ad uno Stato unitario sorto ad appena cinquant'anni di distanza negli stessi luoghi e con personale politico probabilmente ancora in vita. Per quel poco che posso testimoniare, è più facile trovare un'analisi del ruolo degli scienziati italiani a Curtatone e Montanara piuttosto che uno studio comparato delle Accademie italiane nei primi anni del XIX secolo. Eppure tale studio critico sarebbe interessante anche per valutare correttamente, fuori dall'impatto emotivo, pur necessario e da valorizzare, l'origine dei fattori che hanno plasmato gli appuntamenti decisivi a cui l'intera comunità scientifica della Penisola è stata chiamata e a cui aderirà anche con entusiasmo. E forse potremmo capire meglio parte di questo entusiasmo se ci concentrassimo sullo studio delle condizioni di vita di quegli stessi scienziati dieci o venti anni prima. 

\newpage

\nocite{*}

\bibliographystyle{ieeetr}
\bibliography{bibliography.bib} 

\end{document}